# Entangled Photon Correlations Allow a Continuous-Wave Laser Diode to Measure Single Photon, Time-Resolved Fluorescence

*Nathan Harper[1]†, Bryce P. Hickam[1]†, Manni He[1], Scott K. Cushing[1]\**

[1]Department of Chemistry and Chemical Engineering, California Institute of Technology, 1200 E. California Blvd. Pasadena, CA

AUTHOR INFORMATION

†These authors contributed equally to this work

**Corresponding Author**

*Corresponding author: scushing@caltech.edu

# ABSTRACT

Fluorescence lifetime experiments are a standard approach for measuring excited state dynamics and local environment effects. Here, we show that entangled photon pairs produced from a continuous-wave (CW) laser diode can replicate pulsed laser experiments without phase modulation. As a proof of principle, picosecond fluorescence lifetimes of indocyanine green are measured in multiple environments. The use of entangled photons has three unique advantages. First, low power CW laser diodes and entangled photon source design lead to straightforward on-chip integration for a direct path to distributable fluorescence lifetime measurements. Second, the entangled pair wavelength is easily tuned by temperature or electric field, allowing a single source to cover octave bandwidths. Third, femtosecond temporal resolutions can be reached without requiring major advances in source technology or external phase modulation. Entangled photons could therefore provide increased accessibility to time-resolved fluorescence while also opening new scientific avenues in photosensitive and inherently quantum systems.

## TOC GRAPHICS

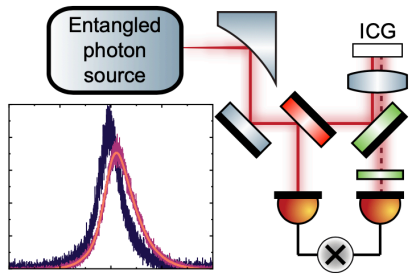

**KEYWORDS** Lifetime, Biosensing, Quantum spectroscopy, Fluorescence, Dynamics

# MAIN TEXT

Time-resolved fluorescence spectroscopy is routinely used in laboratory and clinical settings to measure excited state dynamics, and just as importantly, how changes to the lifetime provide information about the local environment surrounding the fluorophore[1,2]. The advent of inexpensive pulsed laser diodes has led to a proliferation of time-domain fluorescence lifetime techniques in biosensing and microscopy beyond initial physical chemistry applications. Measured lifetimes are intrinsically fluorophore-specific and depend on variations in the surrounding chemical environment, such as temperature[3], pH[4], neighboring fluorophores[5], and the quantum state[6–9]. This specificity, alongside developments in highly fluorescent tagging proteins and molecular markers, has also enabled fluorescence lifetime imaging microscopy (FLIM) to become a ubiquitous technique to visualize and understand complex biological samples[10,11]. Autofluorescence endogenous to cellular environments has permitted label-free FLIM medical imaging applications, including clinical cancer diagnosis[12–14].

The next logical frontier for time-resolved fluorescence is highly distributable, wearable devices[15,16]. At these spatial scales and cost targets, the same pulsed lasers that enabled proliferation of laboratory measurements become a barrier. While on-chip pulsed lasers are being developed, most still require an external power source for the diode or pulse generation[17]. Further, the short on-chip cavity lengths lead to GHz repetition rates that do not give sufficient time for fluorophores to fully relax between pulses[18]. To date, on-chip pulses are mostly in the picosecond range, although recent publications have shown pulses as short as 350 fs[19]. These integrated pulsed lasers typically have a narrow wavelength tuning range, which can make measurements of multiple fluorescent probes or dynamics from multiple local environments difficult[20].

An intriguing alternative to pulsed lasers is to leverage the inherent quantum correlations between entangled photons in a pair. When one high-energy pump photon splits into a pair of two lower-energy entangled photons through spontaneous parametric down-conversion (SPDC), the two single photons are deterministically correlated in time[21]. Thus if one photon is used to excite a fluorophore, coincidence detection of the resulting fluorescence with the other entangled photon provides a measurement of the time-resolved fluorescence[22]. The uncertainty in the temporal width of the photon pair wave packet can range from tens[23–25] to hundreds[21] of femtoseconds, enabling time-resolved measurements comparable to tabletop classical pulsed lasers. It is also key to note that, unlike CW phase modulation spectroscopy[26], a high frequency modulator is not needed for this technique due to its utilization of the inherent correlations between two entangled photons.

Given that classical time-resolved fluorescence instruments are also limited by detector bandwidths and timing circuits, CW entangled photon sources provide comparable experiments, but with a few interesting technological advantages. First, it is straightforward to fabricate on-chip entangled photon sources that require only microwatts of power from a CW laser diode[27,28]. Second, the wavelengths of the SPDC photons can be tuned across the transparency range of the down-conversion crystal, usually from UV to near-IR, by varying the phase matching conditions, either by heating the crystal or tuning the pump wavelength, enabling a range of biosensing and physical chemistry experiments[29]. Finally, femtosecond temporal resolutions can be reached without requiring major advances in external phase modulation or detection technology. Scientifically, given that the entangled photon measurements are natively performed in a single-photon-per-measurement regime, the technique is also interesting for measuring photosensitive samples[30], quantum systems[31], or exploring scaling behaviors of various properties against incidence flux[32,33].

Despite early theoretical interest in utilizing entangled pairs to perform lifetime measurements[22,34,35], an experimental demonstration has not yet been published.[*] We therefore provide a proof of principle experiment by measuring the fluorescence lifetime of indocyanine green (ICG). The correlation time of the entangled photon source is <30 fs[25], but the detector used limits temporal resolution to 50 ps and minimum measurable lifetime to 365 ps[36]. The excited state lifetimes of ICG in three solvents are measured to check the accuracy of the technique against previously published work[37]. Through a statistical analysis we show that, even though limited by the MHz detector maximum count rates and bandwidth, integration times in the minutes range provide sufficient signal for accurate lifetime determination using this configuration. Our results demonstrate that measuring time-resolved fluorescence from entangled photons is possible with properties approaching pulsed lasers, providing the proof of principle for new modalities of compact time-resolved fluorescence spectroscopy and experiments in a single-photon excitation, single-photon detection limit. Periodically poled entangled photon sources routinely approach THz pair generation rates, so commercially available picosecond detectors with GHz count rates can improve these measurements to femtosecond temporal resolution and seconds integration time. Future developments could also perform highly multiplexed measurements using the additional correlated properties of SPDC such as the momentum or angular degree of freedom, as proposed in Ref.[22].

---

[*] At the time of this publication, we became aware of Ted Laurence's unpublished work using a similar method for microscopy.

The experimental configuration for performing fluorescence lifetime measurements with entangled photon pairs is shown in Fig. 1, and a detailed description of all components is provided in the Experimental Methods. Briefly, 2 mW of power from a 402.5 nm CW laser diode pumps a periodically poled, Type-0 SPDC crystal to generate entangled photon pairs spanning 720 nm to 913 nm. Here, the pump laser power is significantly reduced using neutral density filters to avoid saturating the single photon counting detectors. The periodically poled bulk source used here produces free-space pairs, adding down-stream losses due to collimation and fiber coupling. A waveguided equivalent would enable the same studies to be performed on-chip and with μW pump powers.

The energies of the photons in the entangled pair sum to that of the pump laser. For the entangled photon source used, one photon has a wavelength between 720-805 nm with a simultaneously produced conjugate photon between 805-913 nm. The resulting correlation time from this bandwidth is ~30 fs, comparable to the pulse width of state-of-the-art mode locked lasers[38]. The detectors used in this experiment have a larger timing jitter (3.65 ns) than the temporal characteristics of the source and ultimately determine the experimental temporal resolution, as explained later in more detail.

The entangled photon pairs are then deterministically split using a short-pass dichroic filter tuned to 805 nm. The reflected lower energy photon heralds the measurement when it is detected by a single photon counting avalanche detector (SPAD), starting the timer for a time-correlated single photon counting (TCSPC) circuit. The higher energy photon is sent to the sample, where it excites a single molecule that then emits a single fluorescence photon with some delay. The fluorescence photon is collected in an epifluorescence scheme and then detected by another SPAD which stops the timer on the TCSPC board. The time delay between the heralding and fluorescence photon detections is recorded over multiple measurements to create a histogram of such time differences, from which the fluorescence lifetime can be extracted.

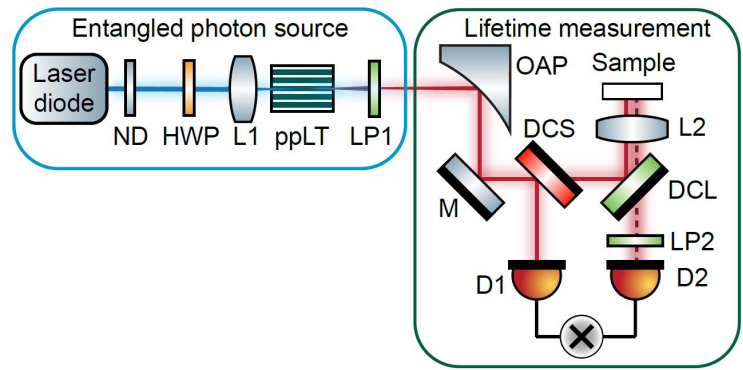

Figure 1: Experimental configuration for measuring fluorescence lifetimes with energy-time entangled photon pairs. The broadband entangled photon source consists of a 402.5 nm laser diode that is focused into a periodically poled lithium tantalate crystal. The resulting entangled photons (solid red lines) are collimated with an off-axis parabolic (OAP) mirror and then separated by wavelength with a dichroic mirror. One photon from the entangled pair is directed to a detector to herald the measurement while the other is focused onto a molecular sample with a high numerical aperture (NA) aspheric lens. The lens collects the backwards emission from the sample's fluorescence (dashed red line) which is then directed through a series of long-pass filters onto a second detector. A coincidence circuit generates histograms of the arrival times between the reference and fluorescence photons. ND: neutral density filters, HWP: half-wave plate, L1: UVFS focusing lens, ppLT: periodically poled lithium tantalate grating, LP1: 500 nm long-pass filter, OAP: off-axis parabolic mirror, M: mirror, DCS: dichroic short-pass mirror (cut-on wavelength 805 nm), DCL: dichroic short-pass mirror (cut-on wavelength 810 nm), L2: high NA aspheric lens, LP2: 808 nm long-pass filter, D1 and D2: multimode fiber coupled single photon avalanche photodiodes connected to a coincidence circuit.

The molecular dye indocyanine green (ICG) is used as a proof of principle since it is commonly utilized to perform fluorescence imaging in biological and clinical settings[39,40]. ICG also has appropriate spectral properties relative to the entangled photon source and a well-characterized lifetime dependence on the solvent environment[37]. Methanol, ethanol, and dimethylsulfoxide (DMSO) were selected as solvents due to their variation of the ICG lifetime by hundreds of picoseconds. The absorption wavelength, emission wavelength, and quantum yield change only slightly across the three solvents (Supporting Information Table 1). The spectrum of the SPDC was tuned through the phase matching temperature to maximize spectral overlap with the sample (Fig. 2a). The resulting ICG fluorescence (Fig 2b) is separated from the excitation beam with a long-pass dichroic mirror and further filtered with an OD6 long-pass interference filter. The instrument response function (IRF, Fig 2b), which characterizes the response that would be measured by a sample with an infinitely short lifetime, was obtained by replacing the sample with a mirror and removing the final long-pass OD6 filter to allow part of the excitation flux to reach the second detector. The IRF spectrum is altered from the excitation spectrum because it has been filtered by the dichroic mirror, but this alteration has minimal impact on the time-domain response due to the minimal wavelength sensitivity of the SPADs used here.

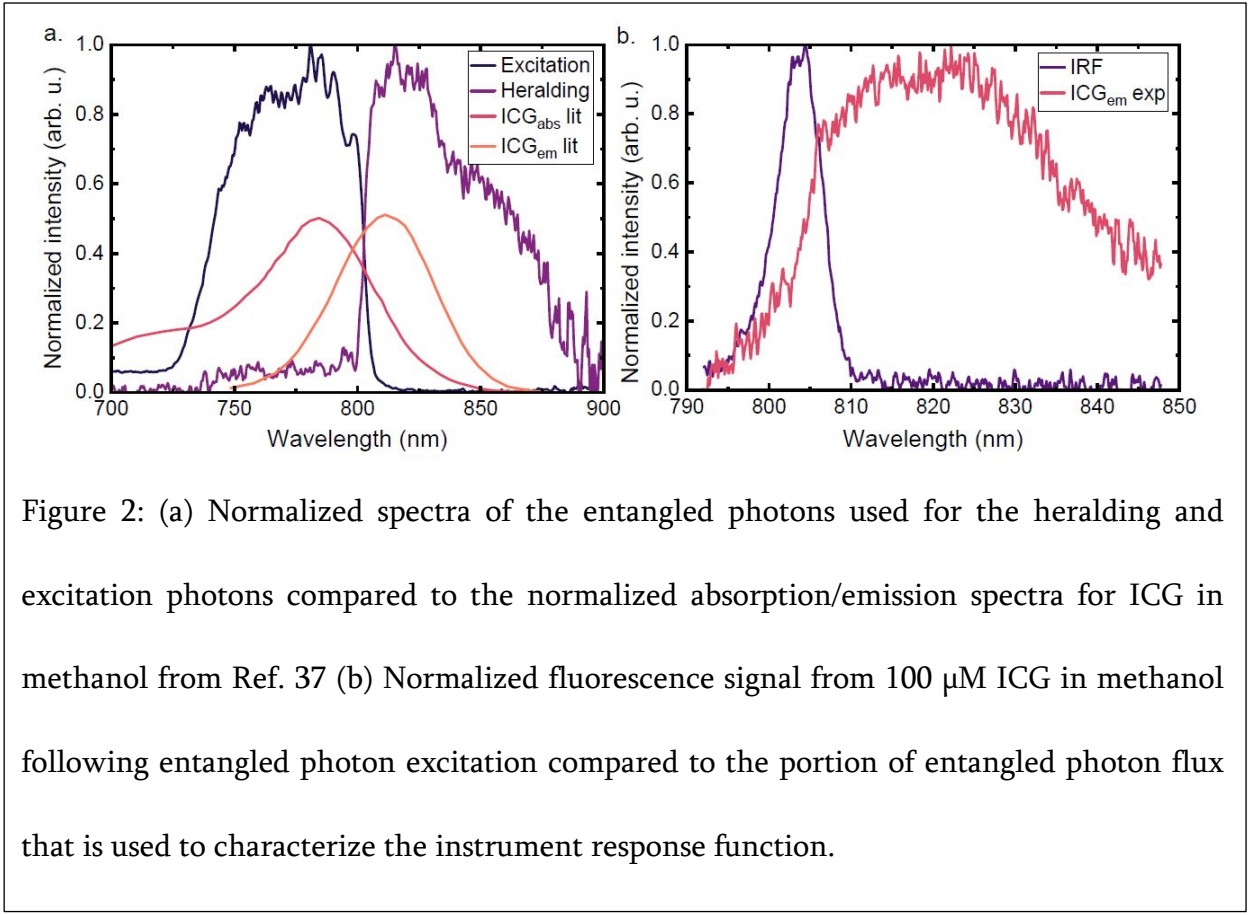

Figure 2: (a) Normalized spectra of the entangled photons used for the heralding and excitation photons compared to the normalized absorption/emission spectra for ICG in methanol from Ref. 37 (b) Normalized fluorescence signal from 100 μM ICG in methanol following entangled photon excitation compared to the portion of entangled photon flux that is used to characterize the instrument response function.

The timing uncertainty broadens the detection of the fluorescence photons by the total IRF. The current temporal resolution of the experiment is limited by the timing uncertainty of the detection electronics, just as with most pulsed laser experiments. The measured histograms R(t) are described by a convolution of the sample's temporal response S(t) with the IRF:

$$R(t) = S(t) * IRF(t) = \int_{-\infty}^{t} S(t')IRF(t-t')dt'$$

After a measurement, the fluorophore lifetime is fit by the sample response function that will give the best match to the measured experimental data. Here, that is a single exponential, but a more complex function could be used if a mixture of lifetimes is present. Therefore, conceptually, the data produced is analogous to TCSPC with a pulsed laser.

From previous theoretical analysis, the minimum fluorescence lifetime that can be measured before deviation in the fit lifetime occurs is approximately 10% of the IRF width[36]. The IRF width in this experiment is approximately 3.65 ns, limiting our instrument to lifetime measurements greater than 365 ps. The counting circuit used here has a minimum time bin width of 4 ps with a 12 ps electrical time resolution, thus the primary factor in the 3.65 ns IRF width is the SPAD timing jitter. It should be noted that detectors with picosecond timing jitter and GHz count rates are commercially available but were not available to our lab at the time of this experiment. The central moment of the IRF drifts over time, in this case with a 50 ps standard deviation over 20 minutes, therefore all lifetimes measured using this approach are reported with this level of uncertainty.

The TCSPC histograms for ICG in three solvents, the IRFs, and the fits of the fluorescence lifetime are shown in Figure 3(abc). The measured TCSPC histograms for ICG in different solvents are shifted and broadened due to the nonzero lifetime. This broadening is more pronounced in DMSO compared to methanol and ethanol because of the longer ICG lifetime in this solvent. The ICG excited state lifetimes from fitting are presented in Table 1 and the exponential decays corresponding to these lifetimes are shown in Figure 3d. The measured lifetimes agree with previously reported values[37] within error. The near-unity $\chi^2$ values and uncorrelated weighted residuals (Supporting Information, Figure 3) suggest the single exponential response fitting is sufficient. The standard deviation of the measurements for all experiments are reported as 50 ps due to detector drift, which is larger than the 3 ps standard errors from the fit. The curves in Figure 3(abc) are normalized and background subtracted for clarity. The raw experimental data plotted on a logarithmic scale is also available in Supporting Information, Figure 3. An experiment where the sample is replaced by a blank cuvette of methanol does not yield a measurable coincidence peak (Supporting Information, Figure 1).

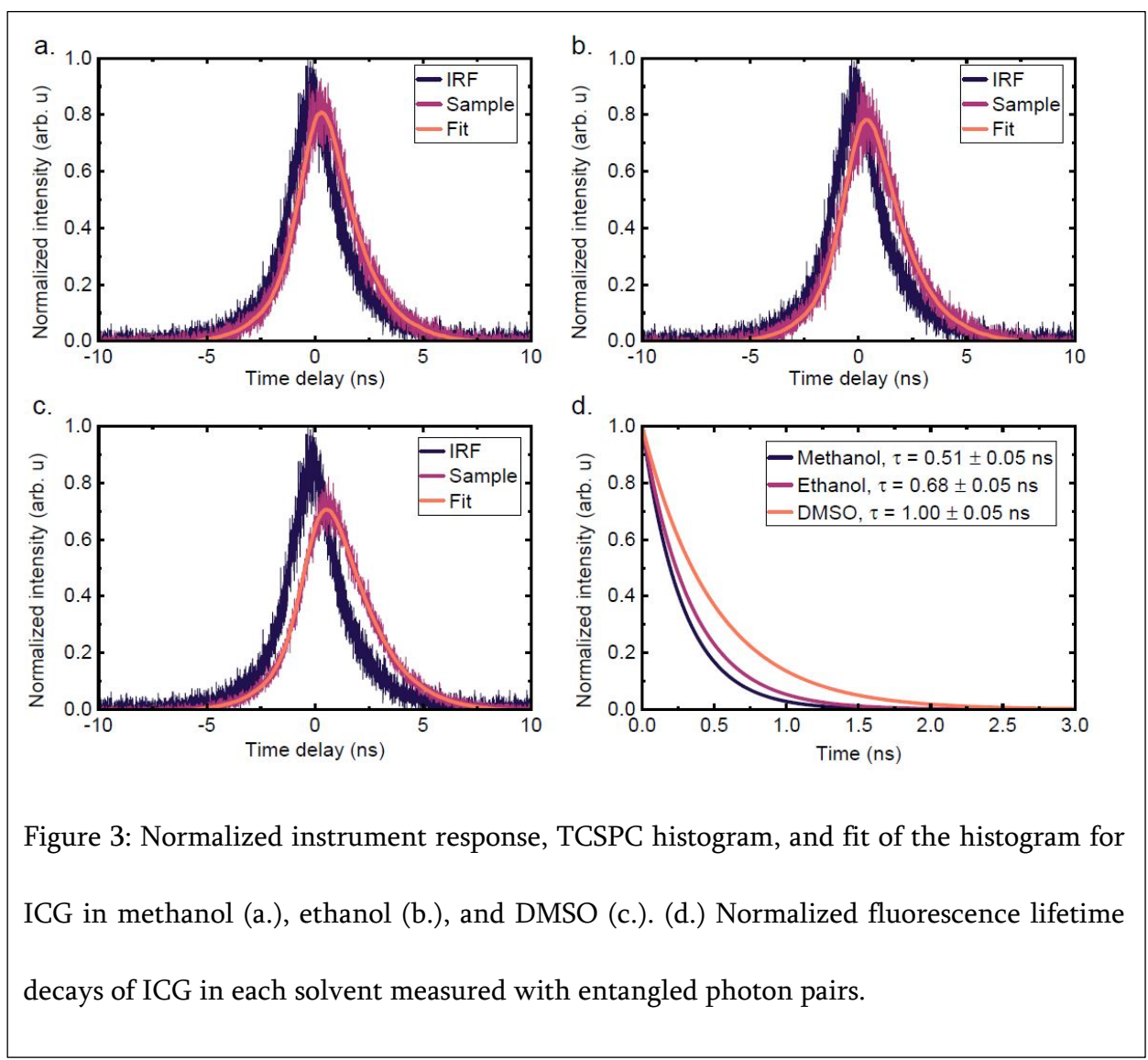

Figure 3: Normalized instrument response, TCSPC histogram, and fit of the histogram for ICG in methanol (a.), ethanol (b.), and DMSO (c.). (d.) Normalized fluorescence lifetime decays of ICG in each solvent measured with entangled photon pairs.

Table 1: Fluorescence lifetimes for ICG in solutions of methanol, ethanol, and DMSO previously characterized in ref.[37] as well as the lifetimes measured in this work and the fitting statistic.

| Solvent | Reported Lifetime (ns) | Measured Lifetime (ns) | $\chi^2$ Statistic |
|---|---|---|---|
| Methanol | 0.51 | 0.56 ± 0.05 | 1.003 |
| Ethanol | 0.62 | 0.68 ± 0.05 | 0.975 |
| DMSO | 0.97 | 1.00 ± 0.05 | 1.019 |

The practicality of entangled photons for fluorescence lifetime measurements should also be quantified. Due to the Poissonian statistics of photon counting, the signal-to-noise ratio is expected to scale according to $\sqrt{N}$, where $N$ is the number of true coincident photon events recorded. The number of coincidences can be increased either by increasing the SPDC flux or by increasing the integration time of each histogram. To verify the noise statistics, the fit procedure was repeated on one-minute subsets of a 14 hour DMSO dataset (Figure 4). As expected, the uncertainty in the fitted lifetime in DMSO scales proportionally to $N^{-0.53\pm0.01}$, close to the expected $N^{-1/2}$ (Figure 4). For each sampling interval, the dataset was binned and a fit was performed on each of the resulting histograms. The resulting standard deviation of the inferred lifetimes was recorded. The increasing uncertainty in the standard deviation at long integration times originates from drift of the counting electronics.

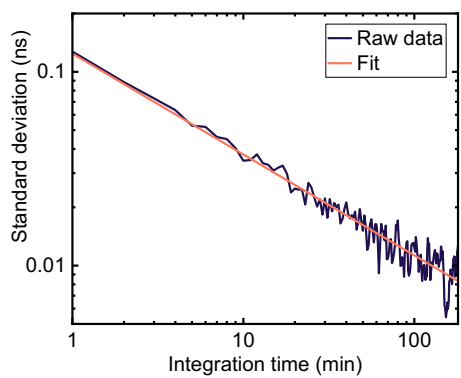

Figure 4: Standard deviation of the fit lifetime with respect to integration time for ICG in DMSO.

Figure 4 suggest that integration times on the order of several minutes would be sufficient for lifetime determination with a 10% uncertainty, based on the detectors used here that saturate at $10^6$ photons/s fluxes. The integration time could be reduced to less than one minute by using higher count rate detectors, such as those that can reach $10^9$ photon/s, given that the pair rate of the experimental entangled source could have been increased by over two orders of magnitude. Theoretically, only ~200 coincidences are necessary for single exponential decay lifetimes to be determined with 10% uncertainty when the IRF is shorter than the molecular lifetime[41], potentially further lowering the experimental integration time to below one second, closer to current microscopic techniques. Using coincidence counting also leads to rejection of background, classical photons on the order of 1,000:1, making the scheme feasible even in sun-lit and room-lit conditions[42].

The proposed technique is therefore aptly timed with the rapidly evolving world of photon counting detector technology[43–45], especially given that THz flux entangled photon sources are already feasible. Figure 4 provides evidence that the use of GHz bandwidth detectors coupled with high efficiency ($10^{-6}$) entangled photon sources would allow for reasonable integration times even with ~µW power CW laser diode lasers. Thus, the technique presented here has multiple interesting outlooks compared to conventional pulsed laser TCSPC or CW modulated fluorescence lifetime measurements, whether it is used for on-chip technology advances, investigation of dynamics in the single-photon excitation limit, or to explore the ability of quantum systems and molecules to preserve incident quantum states of light.

In conclusion, we have demonstrated that molecular excited state lifetimes can be measured using the temporal correlations of entangled photons produced from a CW laser. The solvent-dependent lifetime of indocyanine green was measured in methanol, ethanol, and DMSO in agreement with previous reports using pulsed lasers. The low pump flux needed for SPDC provides a direct route to time-resolved fluorescence measurements using on-chip, CMOS compatible photonics. Multiple waveguides and temperature tuning of phase matching can be used in the future to quickly measure a wide range of fluorophores from a single SPDC source and photonic circuit. Overall, using a CW laser diode as a pump source opens new horizons for performing complex spectroscopic and microscopic studies with quantum light sources.

# EXPERIMENTAL METHODS

*Experimental configuration*

Broadband entangled pairs are generated in a periodically poled crystal whose design and characterization, as well as its use in studying entangled light-matter interactions, has been reported previously[25,46]. In short, photon pairs are generated by Type-0 spontaneous parametric down-conversion in a third-order periodically poled 8% MgO-doped congruent lithium tantalate (CLT) bulk crystal (HC Photonics). The crystal is pumped with a continuous wave diode laser with a maximum power of 400 mW at 402.5 nm and FWHM linewidth of 1 nm (Coherent OBIS). The pump polarization is conditioned with a polarizing beamsplitter (Thorlabs PBSW-405) and half-wave plate (Thorlabs WPH10M-405) before being focused through the SPDC crystal with a 40 cm focal length UV-fused silica lens (Eksma Optics). The temperature of the crystal and subsequent entangled photon bandwidth and spectrum is set and maintained using a chip heater and PID loop with an accuracy of 10mK (Covesion). For measurements of indocyanine green (ICG, Sigma Aldrich) the crystal was heated to 75.0 C to tune the SPDC to a nondegenerate spectrum where the low-wavelength photons overlap with the ICG absorption curve. A series of 500nm OD 4 long-pass (Edmund Optics #84-706) and spatial filters (irises) remove any remaining pump photons and select the entangled photon pairs that are collinear with the pump beam.

Following the filters, the entangled photons are collimated with an off-axis parabolic mirror (Thorlabs) and directed to a short-pass dichroic mirror with a nominal cut-on wavelength of 800 nm (Edmund Optics #69-196). The incidence angle of the dichroic mirror is fine-tuned with a rotation mount (Thorlabs RP005) to increase the cut-on wavelength to 805 nm, the degenerate wavelength of the pairs, as measured with a grating spectrometer (Princeton Instruments IsoPlane SCT-320, 150 gr/mm grating with 800 nm blaze) and electron multiplying intensified CCD (EMICCD, Princeton Instruments PIMAX4). The reflected flux (805 nm – 913 nm) is fiber-coupled into a multimode fiber (105 um core, 0.22 NA, M43L01) using two mirrors and a 4.51 mm focal length aspheric lens (Thorlabs C230TMD-B) and serves as the herald for the experiment. The transmitted flux is directed to a long-pass dichroic mirror with a nominal cut-on wavelength of 800 nm (Edmund Optics #69-883) and serves as the excitation for the sample. The angle of this dichroic mirror is also fine-tuned with a rotation mount (Thorlabs RP005) such that the transmission is >810 nm so that the passband does not overlap with that of the first dichroic mirror. A high-NA 4.51 mm focal length aspheric lens (Thorlabs C230TMD-B) focuses the excitation photons into a 1 mm path length quartz cuvette (Helma/Millipore Sigma) containing ~300 μL of 100 μM ICG in solution. The cuvette is mounted onto a micrometer-driven linear stage to precisely control the position of the sample relative to the focus of the excitation beam. The resulting fluorescence is collected through the long-pass dichroic and coupled into a second SPAD using with two mirrors, a 9.2 mm diameter, 4.51 mm focal length aspheric lens (Thorlabs C230TMD-B), and a multimode fiber (105 um core, 0.22 NA, M43L01). An OD6 808 nm long-pass filter (Semrock LP02-808RE-25) following the dichroic mirror filters any remaining SPDC photons and/or scatter from the fluorescence signal. For the IRF measurements, the sample cuvette was replaced with a mirror, which was positioned so the excitation beam was focused at the surface of the mirror. Differences in pathlength due to the missing glass of the cuvette are estimated to introduce a 5 ps error, which is below the resolution of this experiment. Black cardboard boxes are positioned to block the sample from any pump scatter and minimize background.

Spectral measurements of the heralding and fluorescence photons were performed by collimating the output of the fibers and directing the image onto the spectrometer/EMICCD. The spectrum of the excitation photons was measured similarly by replacing the sample with a multimode fiber with a micrometer-driven x/y translation mount. This same scheme was used with coincidence counting to estimate the number of heralded excitation photons arriving at the sample.

Lifetime measurements were performed by connecting the multimode fibers to SPADs (Laser Components Count) and a coincidence counting circuit (PicoQuant PicoHarp 300). For these experiments, the SPDC between 805 and 912 nm served as the heralding trigger, and the SPAD output was connected to the sync channel of the counting circuit in a forward start-stop mode. The detection events of the fluorescence SPAD served as the stop signal for the experiment, and the SPAD output was connected to the CH1 input of the counting circuit. An electronic delay of 30 ns between the arms was implemented through the counting circuit. This offset is subtracted from the time axis of Figure 3(abc) for clarity but is not removed from the raw data in the supporting information. Neutral density filters were used to decrease pump power and resulting SPDC flux to minimize lost heralding events due to the dead-time of the detectors (43 ns). For lifetime measurements, the number of singles in the heralding arm was approximately $10^5$ counts per second, and the number of singles in the fluorescence arm was approximately 600 counts per second. Data collection consists of repeatedly collecting histograms for 60 seconds, with a series of 10 measurements of the singles rates in each channel measured between each histogram to monitor drift in the alignment. For the methanol, ethanol, and DMSO scans, 840, 840, and 960 histograms were collected, resulting in a total experiment time of 14, 14, and 16 hours, respectively. The instrument response function was measured over 240 coincidence histograms for a total of 6 hours. A custom MATLAB fitting routine is used to determine the characteristic excited state lifetime using the iterative reconvolution approach[47]. In this technique, the single-exponential sample response function is varied so that its convolution with the IRF best matches the measured histograms by minimizing the sum of squared errors.

ASSOCIATED CONTENT

**Supporting Information**

The supporting information file contains:

- Additional experimental data including: select optical properties of ICG in each of the solvents, histograms from an experiment with a blank, probe and fluorescence intensities over the course of the experiment, plots of ICG histograms on a log plot, and residuals from the custom fitting routine.

- Additional details and a description of the fitting routine.

# AUTHOR INFORMATION

## Notes

The authors declare no competing financial interests.

# ACKNOWLEDGMENT

Verification of the lifetimes of the ICG dyes in this study with classical TCSPC were performed at the Caltech Biological Imaging Facility with the help of Giada Spigolon. Verification of the absorption spectra of the dye solutions were performed with the help of Helena Awad. This work was funded by the U.S. Department of Energy (DE-SC0020151). N.H. was supported by the Department of Defense through the National Defense Science and Engineering Graduate Fellowship Program. B.P.H. was supported by an NSF Graduate Research Fellowship (DGE-1745301). Any opinion, findings, and conclusions or recommendations expressed in this material are those of the authors(s) and do not necessarily reflect the views of the National Science Foundation.